\pdfoutput=1

\documentclass[%
 reprint,
 amsmath,amssymb,
 aps,
floatfix,
]{revtex4-2}
\usepackage{algorithm,algpseudocode}
\usepackage{graphicx}
\usepackage{dcolumn}
\usepackage{bm}
\usepackage{comment}
\usepackage[normalem]{ulem}
\usepackage{cancel}

\usepackage{color,xcolor}
\usepackage{eqnarray,amsmath}
\usepackage{xcolor}
\usepackage{array}
\usepackage{booktabs} 
\usepackage{tabularx}

\newcommand{\pp}{{\bm p}}

\renewcommand{\AA}{{\bm A}}

\makeatletter
\def\maketitle{
\@author@finish
\title@column\titleblock@produce
\suppressfloats[t]}
\makeatother

\begin{document}

\preprint{APS/123-QED}

\title{Retrieval of Boost Invariant Symbolic Observables via Feature Importance}

\author{Jose M Munoz}%
\email{jose.munoz25@eia.edu.co}
\affiliation{Department of Physics and Astronomy, Vanderbilt University, Nashville, TN, 37235, USA }

\author{Ilyes Batatia}%
\email{ilyes.batatia@ens-paris-saclay.fr}
\affiliation{Engineering Laboratory, 
  University of Cambridge, 
  Cambridge, CB2 1PZ UK; \\ and 
  ENS Paris-Saclay,
  Université Paris-Saclay, 
  91190 Gif-sur-Yvette, France.}%
 
\author{Christoph Ortner}
\email{ortner@math.ubc.ca}
\affiliation{Department of Mathematics, University of British Columbia, 1984 Mathematics Road, Vancouver, BC, Canada V6T 1Z2.}

 \author{Francesco Romeo}
 \affiliation{Department of Physics and Astronomy, Vanderbilt University, Nashville, TN, 37235, USA.}


\begin{abstract}

Deep learning approaches for jet tagging in high-energy physics are characterized as black boxes that process a large amount of information from which it is difficult to extract key distinctive observables. In this proceeding, we present an alternative to deep learning approaches, Boost Invariant Polynomials~\cite{Munoz:2022gjq}, which enables direct analysis of simple analytic expressions representing the most important features in a given task. Further, we show how this approach provides an extremely low dimensional classifier with a minimum set of features representing 
physically relevant observables and how it consequently speeds up the algorithm execution, with relatively close performance to the algorithm using the full information.
\end{abstract}

\maketitle

\section{Introduction}
\label{sec:Introduction}

Identifying jets originating from hadronically decaying top quarks, known as ``top-tagging'', is a crucial task in the study of high-energy particle physics. It is essential to understand the production and decay mechanisms of the top quark, one of the heaviest known fundamental particles. In recent years, machine learning algorithms have shown promising results in the field of top-tagging, outperforming traditional techniques in terms of efficiency and accuracy. More notably, the field has started to show that the better-performing techniques are the ones that incorporate physical priors such as Lorentz group invariance~\cite{Li:2022xfc}.

In this paper, we apply a method for obtaining the most effective observables for top-tagging in high-energy particle collisions: Boost Invariant Polynomials (BIPs) \cite{Munoz:2022gjq}. We demonstrate how using the BIP technique can capture the intricate correlations found in the jet substructure relying on a polynomial basis to parameterize the distribution of kinematic-based observables which are invariant to rotations, permutations, and Lorentz boosts along the jet axis.

\section{Methodology}\label{sec:methods}

In order to obtain effective observables representative of the signal, we propose to start with sparsified BIP features, fit an ensemble of ML models, like Boosted Decision Trees (BDTs) or Multilayer-Perceptrons (MLPs), for each of them, estimate the feature importance, and select the most important features. Each of those features is given by a simple symbolic expression that can be directly analyzed and interpreted. The details of this strategy are presented throughout the remainder of this section.

\subsubsection{Boost Invariant Polynomials}
We review the Boost Invariant Polynomial (BIP) jet features introduced in \cite{Munoz:2022gjq}. 
We treat a jet as a collection of particles, with each particle specified by its four-momentum $(\pp_i, E_i)$. BIPs are features of the jet, generated via the projection of the momenta to a jet axis $\sum_i \pp_i / N$, where $N$ is the number of particles in the jet. Within this coordinate system, we express the four-momentum of the $i$-th particle in terms of the transverse momentum $p_{\perp,i}$, the transverse energy $E_{\perp,i}$, the angle in an arbitrary reference frame $\varphi_i$ and rapidity $y_i$, i.e.,

\begin{equation}
    \begin{split}
    y_i &= \frac{1}{2}{\rm log}\left(\frac{\delta_1 + E_i + p_{\|i}}{\delta_1 + E_i -  p_{\|,i}} \right),  \\
    E_{\perp,i} &= \sqrt{m_i^2 + p_{\perp,i}^2 },
    \end{split}
\end{equation}

These coordinates are embedded into a polynomial basis and then pooled to obtain permutation-invariant features of the jet, 
\begin{equation}
\label{eq:permutation-basis}
    A_{nlk} = \sum_{i=1}^N B_n(\tilde{p}_{\perp,i}) \log(1 + E_{\perp,i})
        e^{\mathrm{i} l \varphi_i} e^{- \lambda k y_i},
\end{equation}
where $B_n$ are the Bessel polynomials and $\tilde{p}_{\perp,i}$ refers to a custom transformation of the transverse momentum of the form:
\begin{equation} \label{eq:transform_p}
    \tilde{p}_{\perp,i} = A \log\Big( { \frac{p_{\perp,i}}{\sum_i p_{\perp,i}}} + \delta_2 \Big) + B.
\end{equation}

The BIP features are obtained by taking $\nu$ correlations, 
\begin{equation}
    \label{eq:product-basis}
    \begin{split}
    &\AA_{\boldsymbol{nlk}} = \prod_{t=1}^{\nu} A_{n_t l_{t}k_{t}}, \qquad \\
    &\text{with   }
    \boldsymbol{nlk} = (n_{1}l_{1}k_{1},\ ...,\ n_{\nu}l_{\nu}k_{\nu}) ,\quad \nu > 0, \\ 
    & \text{and } \sum_t l_t = \sum_t m_t = 0,
    \end{split}
\end{equation}
The final condition guarantees that the features $\AA_{\boldsymbol{nlm}}$ are invariant under rotations of the jet about its axis and invariant under boosts along the jet axis. 


A sparse subset of the BIP basis is then selected as an input layer to a classifier. To that end we introduce a maximum correlation order $\bar \nu$ and a level $\Gamma$ and retain only those features $\AA_{\bm nlk}$ for which 
\begin{equation}
    \label{eq:totaldegree}
    \sum_{t = 1}^\nu |l_t| + |k_t| + n_t \leq \Gamma
    \quad \text{and} \quad \nu \leq \bar\nu.
\end{equation}
%
It was shown in~\cite{Munoz:2022gjq} that increasing the values of these parameters generates a more expressive basis but also makes fitting a shallow model more difficult.

\subsubsection{Ensembles of ML models}

In \cite{Munoz:2022gjq} we demonstrated that the BIPs feature provides an excellent representation of a jet via fitting out-of-box classifiers.
However, a considerable improvement can be expected from generating an ensemble of diverse classifiers and taking their arithmetic average as the final discriminator. We achieve this by fitting a collection of classifiers such as simple Multilayer-Perceptron (MLP), different Gradient boosting methods such as CatBoost and XGBoost, and more traditional ones such as Random Forest and Logistic Regressions.

For each of these classifiers, a selection is set up: in each of the rounds, the 3 best performing algorithms are hyper-tuned using Optuna~\cite{akiba2019optuna}, and this process is repeated twice. Although this may appear computationally demanding, due to the low computational cost and low dimensionality of BIPs, the process is in fact considerably faster and more computationally efficient than that of training state-of-the-art Transformer architectures.

\subsubsection{Estimation of the Feature Importance}
One of the most concerning issues in the usage of ML in physics applications is the difficulty of building mathematically accessible observables from those models.
To overcome this challenge, we propose the usage of SHapley Additive exPlanations (SHAP) as a method for estimating feature importance in machine learning models \cite{10.5555/3295222.3295230}. It provides a flexible framework for interpreting the predictions of any black-box model, allowing us to understand the impact of each feature on the model output.

At its core, SHAP values are based on the concept of Shapley values, which were originally developed in game theory to allocate the payoff of a game among its players. In the context of feature importance estimation, the Shapley value of a feature represents its contribution to the model output on average over all possible combinations of features. The importance scores are computed using a set of reference samples and a set of perturbed samples. The reference samples are used as a baseline to compare the model output to, while the perturbed samples are used to simulate the effect of adding or removing each feature. By computing the difference between the model output for the perturbed and reference samples, the contribution of each feature to the model output is estimated.

A key strength of SHAP values is that they can be applied to complex interactions between features. Unlike simpler methods such as permutation importance or partial dependence plots, SHAP values can capture non-linear and non-monotonic relationships between features.
Moreover, they can be used to generate global and local explanations of model predictions. Global explanations provide an overview of the relative importance of each feature in the model, while local explanations provide an explanation of a specific prediction, highlighting the features that had the greatest impact on that prediction.

One of the most important assumptions made by the SHAP method is the independence between features, which often fails for both high and low-level features. However, by construction, our BIP features are linearly independent, allowing a safe usage of the importance scores. 

Another challenge when inferring the most important features is the discrepancy between the predictions made by the different architectures (MLP, SGBoost, etc). In order to minimize this effect, we estimate the SHAP values for an ensemble of our 5 best-performing models on the validation dataset.

Obtaining the indices for the most important features yields little information about the actual physics underlying the features. In order to extract the actual observables, one has to ``back-propagate'' the index to the corresponding polynomial expression. For each one of those features selected in the tensorial basis $\AA_{\boldsymbol{nlk}}$, one obtains the respective tuple of $(n, l, k)$ multi-indices that generates that feature, from which the corresponding symbolic polynomial expression is readily obtained.

\section{Results}

\subsection*{Top Tagging}
This study uses the top tagging dataset proposed in~\cite{Kasieczka:2019dbj}, which includes 1.2 million training data points, 400,000 validation data points, and 400,000 test data points. Each data point represents a jet that originates from a top quark (the signal) or from a light quark/gluon (the background), and contains the kinematic information of up to 200 particles with an average of 30 particles for each jet. This kinematic information, i.e. the four-momenta of the jet constituents, forms the input variables (full basis) for the reference model. The jets were generated at 14 TeV and underwent a simplified simulation of the ATLAS detector in Delphes \cite{deFavereau:2013fsa}. All jets are selected requiring $p_T \in [550,650]$ GeV, $|\eta | < 0.2$, and $\Delta R = 0.8$ of the jets axis. 

\begin{figure}[!h]
    \centering
    \includegraphics[width=0.4\textwidth]{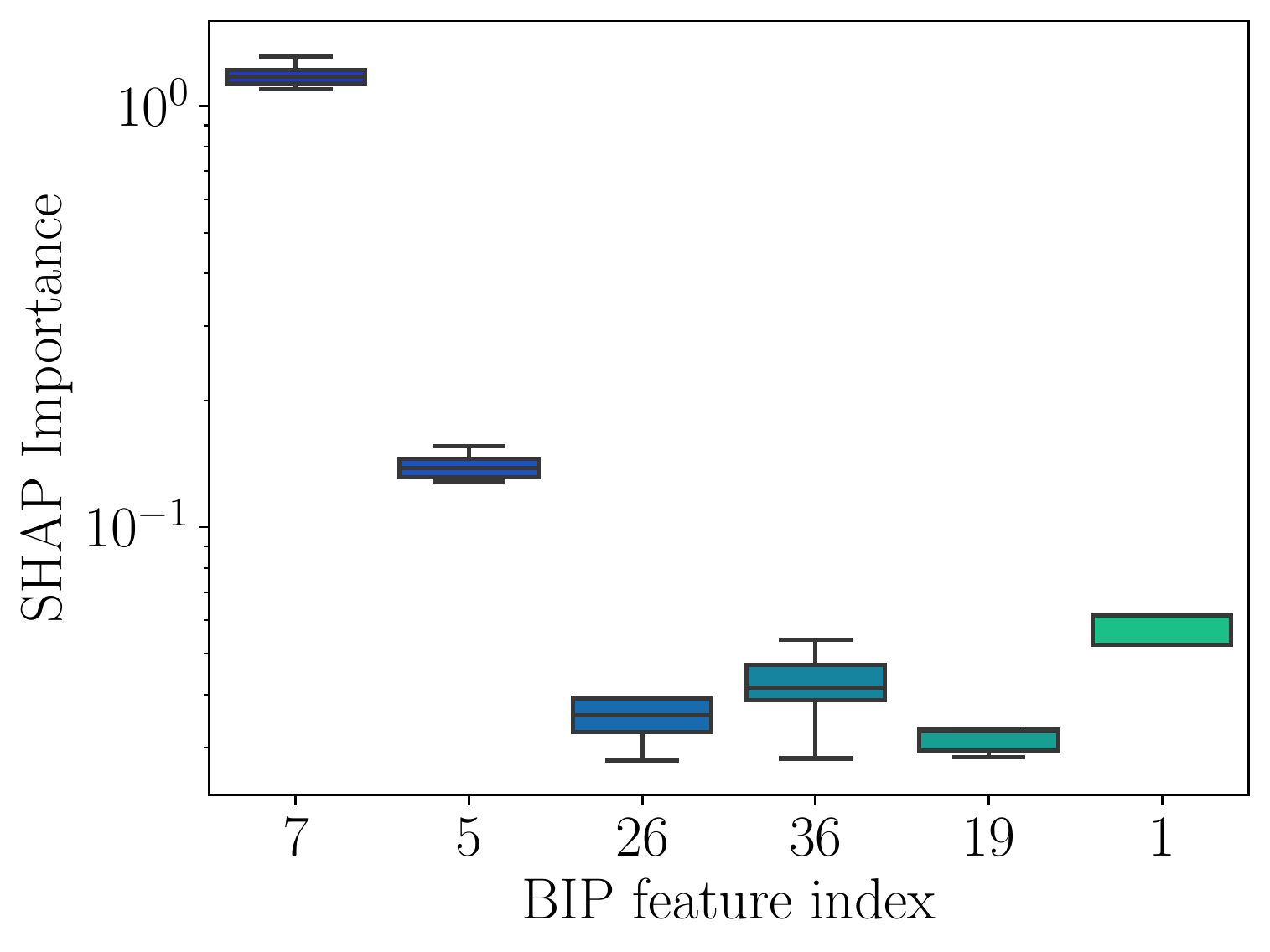}
    \caption{\centering Box-plots distribution of the global SHAP importances for the six highest ranked features in the BIP($\nu=2$, $\Gamma=5$) basis. }
    \label{fig:shap_importance}
\end{figure}

The first step is to construct the BIP features  for the training, validation, and testing sets using a relatively small number of features, 52, resulting from the hyperparameter choices $\bar \nu=2$ and $\Gamma=5$.
In the five different folds used for the training data, we found consistent feature importance, both for the signal and the background, as shown in Fig.~\ref{fig:shap_importance} where the distribution of the feature importance for the different folds of the training data is shown with the box plots. The black lines, representing the extrema importance scores, show that for the there is a clear dominance for the most important features.

\begin{figure}[!h]
    \centering
    \includegraphics[width=0.45\textwidth]{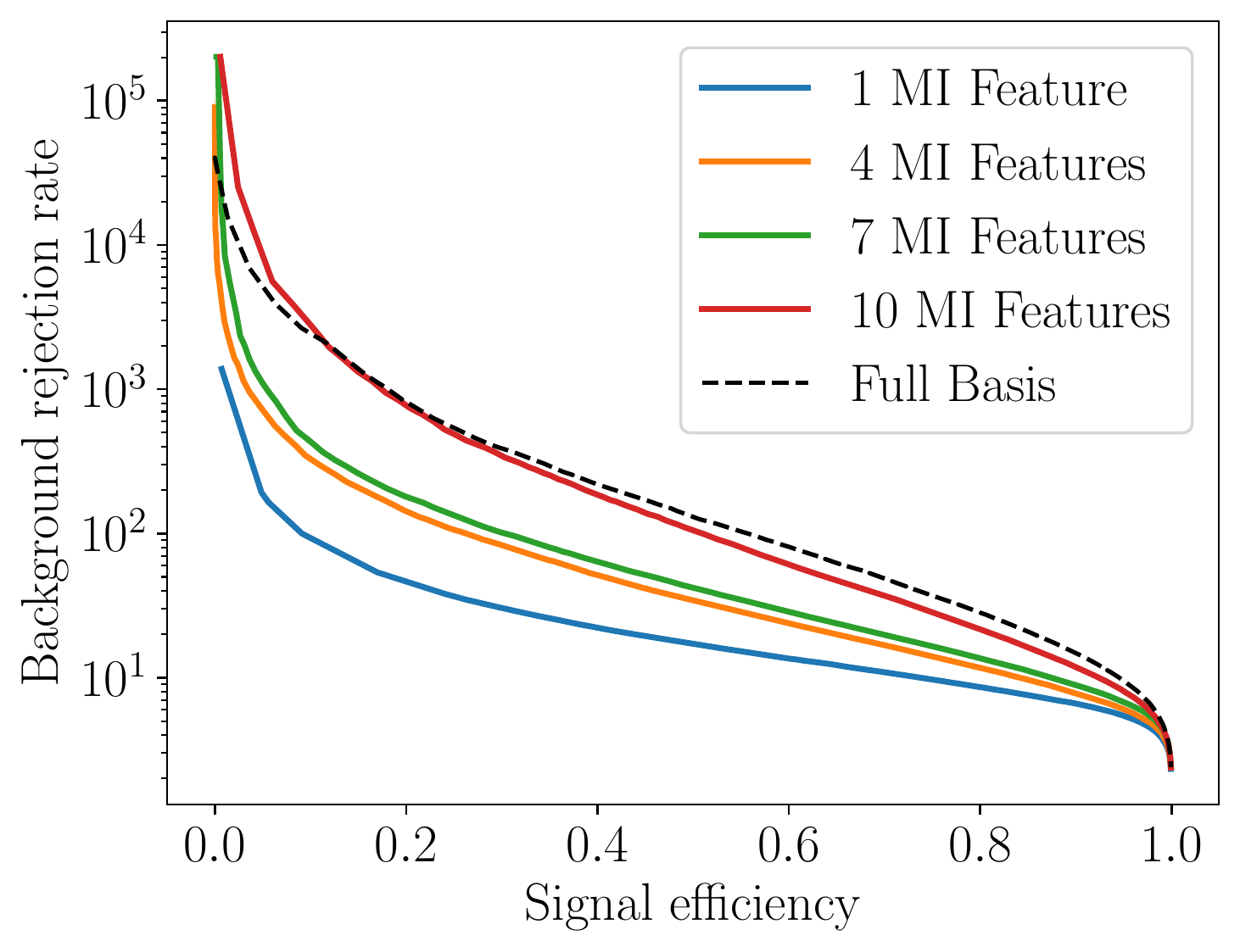}
    \caption{Signal Efficiency versus Background Rejection Rate for the classification performed on a different number of the most important (MI) BIP features.}
    \label{fig:rej_rate}
\end{figure}

We train a Boosted Gradient Classifier with the full BIP($\bar\nu = 2$, $\Gamma = 5$) basis and take it as a baseline for comparison. We then order the most important features for the BIP basis by decreasing SHAP value, and thus define 10 new models to train with the same model parameters but add to each model the the next BIP feature based on SHAP importance value. The first model will use only the feature with the highest SHAP value representing the most important (MI) feature, the second model the 2 features with the 2 MI features and so forth, as it is performed in forward-selection methods \cite{Das:2022cjl}. It is remarkable that using only the single most important feature, which is computed via a simple symbolic equation obtained by the BIP feature representation we propose, we obtain an Area under the curve (AUC) of 0.929 and an accuracy of 0.890. Increasing the number of features correspondingly increases the performance up to the case in which using the best 10 BIP features we reach a similar performance to baseline training, as shown in Fig.~\ref{fig:rej_rate}.

Next, we extract the symbolic expressions for the six features with the highest SHAP values and write them out in Equation~\eqref{eq:symexpr}.

\begin{widetext}
\begin{subequations} \label{eq:symexpr}
\begin{align} 
    \label{eq:sym7}
 \AA^{(7)} &= A_{0,0,0} \cdot A_{0,0,0} &&\hspace{-1.6cm}=\Big[\sum_{i=1}^N \log(1 + E_{\perp,i})\Big]^2 \\ 
    \label{eq:sym5}
    \AA^{(5)} &= A_{4,0,0} &&\hspace{-1.6cm}= \sum_{i=1}^N B_4(\tilde{p}_{\perp,i}) \log(1 + E_{\perp,i}), \\
    \label{eq:sym26}
    \AA^{(26)} &= A_{0,1,0}\cdot A_{3, -1, 0}
    &&\hspace{-1.6cm}=\Big[\sum_{i=1}^N 
        e^{\mathrm{i} \varphi_i} \Big]\cdot \Big[\sum_{i=1}^N B_3(\tilde{p}_{\perp,i}) \log(1 + E_{\perp,i})
        e^{-\mathrm{i} \varphi_i}\Big], \\ 
        \label{eq:sym36}
    \AA^{(36)} &= A_{0,-1, 1}\cdot A_{1,1, -1}
    &&\hspace{-1.6cm}=\Big[\sum_{i=1}^N \log(1 + E_{\perp,i})
        e^{-\mathrm{i} \varphi_i- \lambda y_i}\Big] \cdot \Big[\sum_{i=1}^N (\tilde{p}_{\perp,i}+1) \log(1 + E_{\perp,i})
        e^{\mathrm{i} \varphi_i + \lambda y_i}\Big], \\ 
        \label{eq:sym19}
    \AA^{(19)} &= A_{0,0,-1}\cdot A_{2, 0, 1}
    &&\hspace{-1.6cm}=\Big[\sum_{i=1}^N 
        e^{\lambda y_i} \Big]\cdot \Big[\sum_{i=1}^N B_2(\tilde{p}_{\perp,i}) \log(1 + E_{\perp,i})
        e^{-\lambda y_i}\Big], \\ 
        \label{eq:sym1}
    \AA^{(1)} &= A_{0,0,0} &&\hspace{-1.6cm}=\sum_{i=1}^N \log(1 + E_{\perp,i})
\end{align} 
\end{subequations}
\end{widetext}

\section*{Interpretation}
We discuss the impact on the top-quark tagging task of having the BIP features defined above and ranked by best SHAP value.

\begin{figure}[!h]
    \centering
    \includegraphics[scale=0.4]{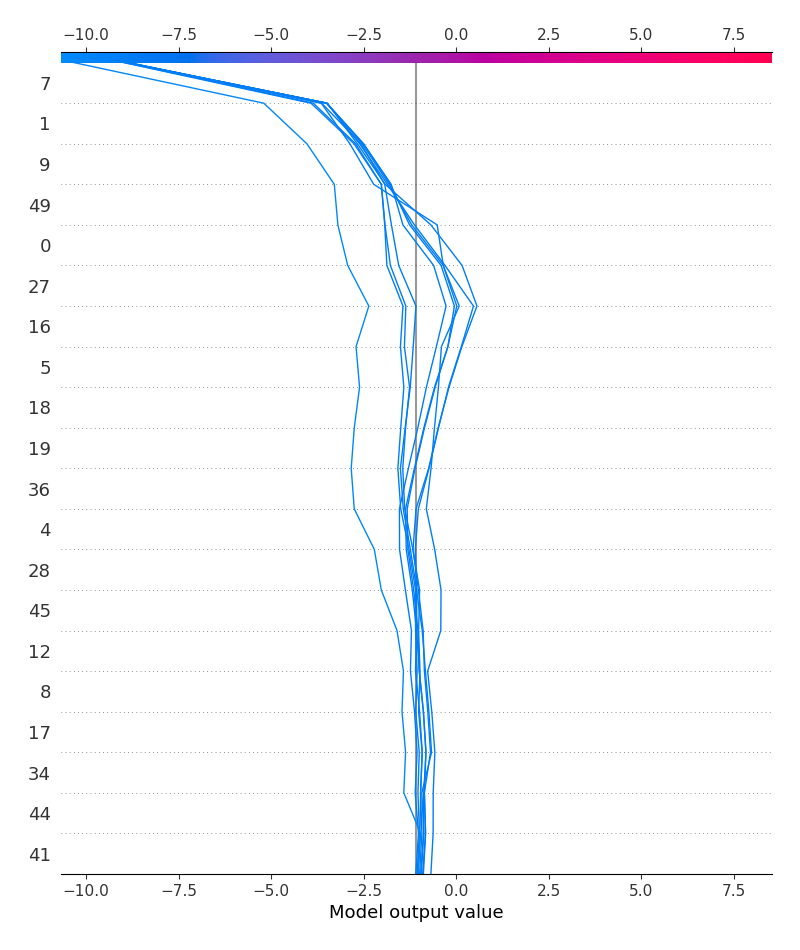}
    \caption{Distribution of the feature importance for the best decisions for the background jets.}
    \label{fig:learner_fold_3_shap_class_0_best_decisions}
\end{figure}

\begin{figure}[!h]
    \centering
    \includegraphics[scale=0.4]{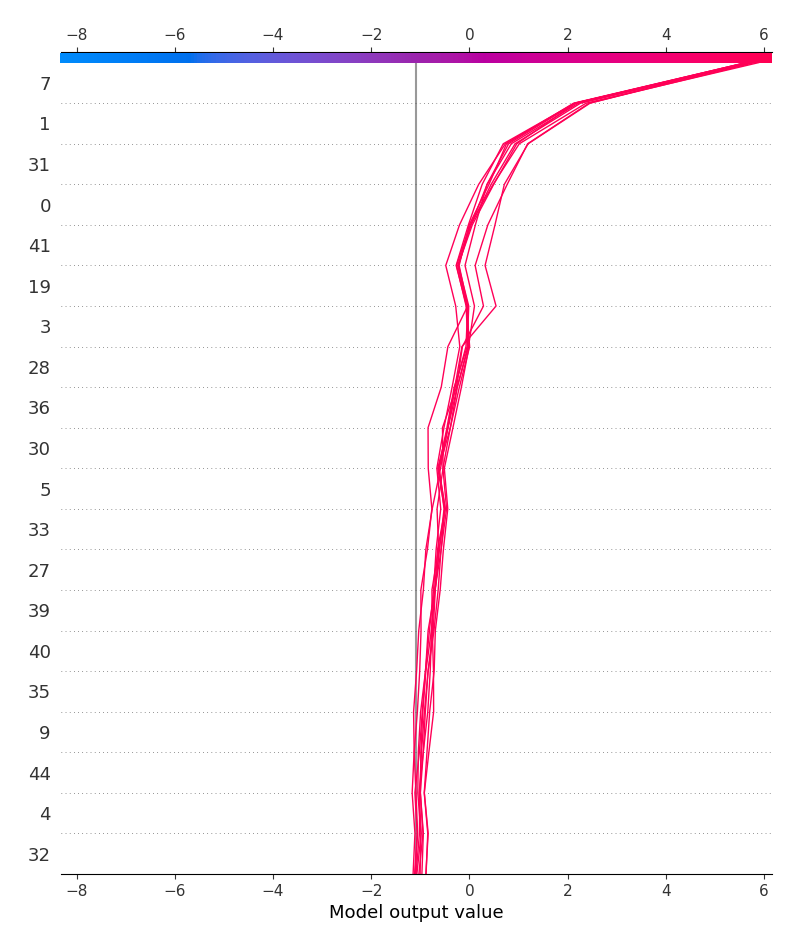}
    \caption{Distribution of the feature importance for the best tagging decisions for the top jets.}
    \label{fig:learner_fold_3_shap_class_1_best_decisions}
\end{figure}

The performance of the machine learning model that uses only the highest BIP feature, defined in \eqref{eq:sym7} has degraded performance compared to the baseline case. This is expected because the simplest BIP expression in \eqref{eq:sym7} does not include information about the jet constituent $p_T$, which is a highly discriminating observable and is used in the baseline training. At the same time, as we increase the number of BIP features we enhance the amount of discriminating information and we consequently improve the tagging performance, up to converging to the performance of the baseline training with 10 BIP features. The distribution of the feature importance is shown in Figs.~\ref{fig:learner_fold_3_shap_class_0_best_decisions} and \ref{fig:learner_fold_3_shap_class_1_best_decisions},
which illustrates how the significance of the features rapidly converges, indicating that the additional information provided by the full basis is mostly redundant for distinguishing between signal and background. 
The BIP features are capable of providing high discrimination because they are able to intercept physically relevant aspects. For example, the high multiplicity of jet constituents in hadronically decaying top quarks in the boosted topology, which consists of 3 quarks, compared to the background jets, which consist of 1 gluon or 1 quark is reflected in the expressions \eqref{eq:symexpr} which are defined as sums of the signal and background jet constituents.

\section*{Conclusions and Outlooks}

We have shown that it is possible to build an extremely small subset of rotation-, permutation-, and boost-invariant polynomial features for performing jet tagging by retaining only a small number of features with the highest SHAP importance score until the performance with the full-basis is matched.
For a specific application case considering the signal given by hadronically decaying boosted top quarks and the background given by non-top quarks and gluons in a restricted kinematic region, and relying just on kinematic information, we demonstrate that the BIP feature definition can obtain comparable performance to the full basis with just 10 variables over 62 variables in the full basis. 
These features are mapped to the BIPs tensorial indices to infer the symbolic expressions that generate them and render a well-defined and interpretable set of observables that can provide additional insight into the underlying physical process, a potential advantage compared with other more black-box deep learning approaches to jet tagging.
In addition to significantly reducing the required set of input variables to achieve a high level of discrimination, this approach allows for a much faster inference process, which is a desirable characteristic of any online application for jet tagging, such as runtime event triggering of events in high-energy physics.
This procedure can be extrapolated to a variety of jet-representation tasks, not only related to classification problems but also to regression problems, such as inferring the mass or other properties of a given jet.
The results achieved here represent a consistent proof-of-concept but in order to establish its effectiveness 
for more realistic applications in high-energy experiments, it must be tested on a greater variety of application scenarios.
\nocite{*}
\bibliography{main/main}

\end{document}